\documentclass{aastex62}
\usepackage{multirow}
\usepackage{makecell}
\usepackage{upgreek}

\def\lax{{$\mathrel{\hbox{\rlap{\hbox{\lower4pt\hbox{$\sim$}}}\hbox{$<$}}}$}}
\def\gax{{$\mathrel{\hbox{\rlap{\hbox{\lower4pt\hbox{$\sim$}}}\hbox{$>$}}}$}}

\begin{document}

\title{Dependence of Spiral Arms Pitch Angle on Wavelength as a Test 
of Density Wave Theory}

\author[0000-0002-3462-4175]{Si-Yue Yu}
\affiliation{Kavli Institute for Astronomy and Astrophysics, Peking University, Beijing 100871, China}
\affiliation{Department of Astronomy, School of Physics, Peking University, Beijing 100871, China}

\author[0000-0001-6947-5846]{Luis C. Ho}
\affiliation{Kavli Institute for Astronomy and Astrophysics, Peking University, Beijing 100871, China}
\affiliation{Department of Astronomy, School of Physics, Peking University, Beijing 100871, China}

\begin{abstract}
Large-scale galactic shocks, predicted by density wave theory, trigger 
star formation (SF-arms) downstream from the potential of the oldest stars (P-arms), resulting in 
a color jump from red to blue across spiral arms in the direction of rotation, 
while aging of 
these newly formed young stars induces the opposite but coexisting classic age gradient further downstream from the SF-arms.
As the techniques for measuring pitch angle are intensity-weighted,
they trace both the SF-arms and P-arms and are not sensitive to the classic age gradient.
Consequently, the measured pitch angle of spiral arms should be systematically smaller in bluer bandpasses
compared to redder bandpasses.
We test these predictions using a comprehensive sample of 
high-quality optical ({\it BVRI}) images of bright, nearby spiral galaxies 
acquired as part of the Carnegie-Irvine Galaxy Survey, supplemented by 
{\it Spitzer}\ 3.6 $\micron$ data to probe evolved stars and {\it GALEX}\ 
ultraviolet images to trace recent star formation.  We apply one-dimensional 
and two-dimensional techniques to measure the pitch angle of spiral arms, 
paying close attention to adopt consistent procedures across the different 
bandpasses to minimize error and systematic bias. 
We find that the pitch angle of spiral arms decreases mildly 
but statistically significantly from the reddest to the bluest bandpass, 
demonstrating conclusively that young stars trace tighter spiral arms than old 
stars. Furthermore, the correlation between the pitch angle of blue and red 
bandpasses is non-linear, such that the absolute value of pitch angle 
offset increases with increasing pitch angle.  Both effects can be naturally 
explained in the context of the density wave theory for spiral structure.
\end{abstract}


\section{Introduction}

Spiral structure is the most striking feature of disk galaxies, but its
physical origin is still debated.  The density wave theory proposed by
\cite{LinShu64}, perhaps the most successful framework proposed for spiral
structure, envisages a quasi-stationary wave pattern rotating around the
galactic center at a constant angular speed.  The spiral potential generated
by the oldest stars induces large-scale galactic shocks on the gas, triggering
gravitational collapse and then enhanced star formation \citep{Roberts1969}. 

The enhanced star formation caused by density waves can lead to 
two opposite but coexisting color gradients across spiral arms.
Firstly, gas clouds get
shocked upstream from the minima of the spiral potential (P-arms) and take a
finite timescale to form arms with enhanced star formation (SF-arms)
downstream from the P-arms \citep{Gittins2004}. As the spiral potential is 
generated by the oldest stars, a red-to-blue color 
gradient occurs in the direction of rotation for trailing spirals.
The second color gradient---the classic age color gradient---comes from
the aging of newly formed young stars of the SF-arms.
Inside the corotation radius, the newly formed stars drift differentially out of their birth site and age meanwhile.
This drift causes an age color gradient from blue to 
red in the direction of rotation.  
In summary, there is a spatial ordering of different tracers across spiral arms in the direction of 
rotation: the oldest population, the youngest stars, and a gradually aging population.
Due to the requirement that arms vanish at corotation radius (CR), the
azimuthal offsets among them decrease
with increasing radius inside 
CR and increase thereafter (or beyond CR), 
implying that the pitch angle of these tracers follows: pitch angle of 
P-arms $>$ pitch angle of SF-arms $>$ pitch angle of aging red arms.
This picture, though, is less clear-cut in reality. Smearing-out effects between 
the gas and stars, due to the continuous formation of stars behind the shock and the
tendency for newly formed stars to fall to smaller galactocentric radii due to loss of 
angular momentum, reduce the asymmetry of the classic color gradient anticipated 
by theory \citep{YuanGrosbol1981}. Furthermore, the stronger shocks at smaller radii 
increase the inward streaming motions there, such that the azimuthal offsets narrow, and mimic
a reduction in the pattern speed relative to that measured at larger radii \citep{Martinez2009b}.
Evidence for the classic age color gradient has been found
by \cite{Gonzalez1996}, \cite{Martinez2009a}, and \cite{Martinez2011}
by using a photometric index, $Q$({\it r}\,{\it J}\,{\it g}\,{\it i}), that effectively 
traces the 
gradient from young to relative old stars, 
while contradictory results have also been reported \citep{Schweizer1976, 
Talbot1979, Foyle2011}.  
One-dimensional (1D) or two-dimensional (2D) Fourier transformation are widely 
adopted techniques to measure pitch angle of spiral arms. Because these two 
techniques implicitly use the intensity as weighting when calculating the
Fourier components, they will trace the centroid of spiral arms near their peak, 
thereby measuring the pitch angle of P-arms in red-band and SF-arms in 
blue-band images. Therefore, in observational terms, spiral arms should 
have smaller pitch angles in the blue than in the red if 
they are density wave modes. A number of studies 
indeed have found evidence of spiral arms being tighter in bluer than in 
redder bands \citep{Grosbol1998,Martinez2012,Martinez2013,Martinez2014}.
\cite{Martinez2013}, in particular, reported a median difference in pitch angle
of approximately $-1\degr$ between the {\it g}\ and {\it J}\ band for a sample
of 11 objects.  By contrast, neither \cite{Seigar2006} nor \cite{Davis2012}
found discernible variation in pitch angle with waveband.  \cite{PI2016},
extending the range of wavelengths to the far-ultraviolet (FUV), even came to
the {\it opposite}\ conclusion: spirals are looser in the ultraviolet and grow
tighter toward the red.  A major source of difficulty with these analyses is
that the pitch angle variations are expected to be small, and hence
measurement uncertainties and systematic effects become very important, both
within each band and across different bands.   If we wish to compare pitch
angles from different images, great care must be taken, when performing
the measurements, to choose strictly consistent galaxy parameters, such as
galaxy center, ellipticity, position angle, and radius range.

No color dependence in arm tightness is expected for the transient but 
recurrent spiral structures 
presented in $N$-body simulations \citep[e.g.,][] {Sellwood2011, 
Baba2013, Onghia2013}
obeying swing amplification theory 
\citep{Toomre1981}. Thus, measuring the pitch angle as a function of 
wavelength provides a simple test for theories of spiral structure.  

\cite{Yu2018} recently measured robust pitch angles for a large sample of
spiral galaxies drawn from the Carnegie-Irvine Galaxy Survey \citep[CGS;][]{
Ho2011}, using two independent methods based on 1D and
2D Fourier decomposition of optical ({\it BVRI}) images.  
Here, we extend the analysis of a subset of this sample to an even wider 
wavelength baseline, by making use of 3.6 $\mu$m IRAC images from the {\it 
Spitzer}\ Survey of Stellar Structure in Galaxies \citep[S$^4$G;][]{Sheth2010},
which are an ideal extinction-free tracer of the stellar mass distribution in
galaxies \citep{Meidt2014}.  At the opposite extreme, we further include
FUV and near-ultraviolet (NUV) images from the {\it Galaxy Evolution Explorer}\
\citep[{\it GALEX};][]{Martin2005} to sample the most recently formed stars.
Together with the CGS data, our analysis covers a total of seven bandpasses
from $\sim 1500$ \AA\ to 3.6 $\mu$m, offering the most comprehensive view to
date of the wavelength dependence of spiral arm pitch angle.

\section{Data}

The CGS is a statistically complete optical ({\it BVRI}) imaging survey of 605
bright ($B_T \le 12.9$ mag), nearby (median $D_L=24.9$ Mpc), southern
($\delta < 0\degr$) galaxies. The overall quality of the images is quite high,
having a median seeing of $\sim1\arcsec$ and a limiting surface brightness of
$\sim$25.3 mag~arcsec$^{-2}$ in the {\it I}\ band. \cite{Yu2018} successfully 
and systematically measured the pitch angles of spiral arms for 172 CGS galaxies
in {\it BVRI}, using both 1D and 2D Fourier decomposition.  We crossmatch the 
spiral galaxies with pitch angle measurement in \cite{Yu2018} with S$^4$G 
and {\it GALEX}.  We excluded galaxies with spiral arms in {\it I}-band
images that are too flocculent or irregular to measure reliable pitch angle 
\citep{Yu2018}. The 3.6 $\micron$ images reach a limiting surface brightness 
of $\sim27$ mag arcsec$^{-2}$ and thus are quite deep, but the quality of the 
NUV/FUV images is spotty. Images from the {\it GALEX}\ All-Sky Imaging Survey 
(AIS) are rather shallow, and we exclude them if the spiral arms are 
indistinguishable. We finally have 82 $3.6\,\micron$ images, 80 NUV images 
[44, 6, 10, and 20 of them are from {\it GALEX} AIS, Medium Imaging Survey 
(MIS), Nearby Galaxy Survey (NGS), and Guest Investigator Data (GI), 
respectively] and 71 FUV images (38, 6, 10, and 17 of them from {\it GALEX}\
AIS, MIS, NGS, and GI, respectively).

An accurate determination of the sky projection parameters--- ellipticity ($e$)
and position angle (PA)---for the galaxies is essential for the study of spiral
arms. \cite{Yu2018} carefully scrutinize the projection parameters from
different sources to choose the optimal values.  For the current, demanding
application in hand, the problem is even more acute because we need to compare
potentially small differences in pitch angle across multiband images with
inhomogeneous resolution and signal-to-noise ratio.  It is also particularly
challenging that within any given galaxy the pitch angle variations with
radius can exceed 20\% \citep{Savchenko2013}.  Thus, it is crucial to keep
fixed the $e$, PA, and centroid of the galaxy, as well as the radial range 
occupied by spiral arms, when calculating the pitch angle for images in 
different wavebands.  This procedure was followed strictly
for the {\it BVRI}\ CGS images \citep{Yu2018}, and here we extend it to the
IRAC and {\it GALEX}\ images, adopting, as reference, the projection parameters
from the CGS $I$-band image (Table 1).  The absolute astrometry for the IRAC
and {\it GALEX}\ images is determined according to the world coordinate system
of the galactic center of the $I$-band images \citep{Li2011}.\footnote{The
world coordinate system for a few CGS images is either unavailable or has a
slight orientation offset.  For these particular data, we aligned the
IRAC/{\it GALEX}\ images with the CGS images by visual inspection.}
 
Prior to performing Fourier decomposition to measure the pitch angle, the
images need to be background-subtracted, and bright foreground stars must be
removed.  These procedures have already been performed for the CGS images 
\citep{Ho2011, Li2011}, and we apply them in the same manner to the
IRAC and {\it GALEX}\ images.

\begin{figure}
\figurenum{1}
\centering
\includegraphics[width=14cm]{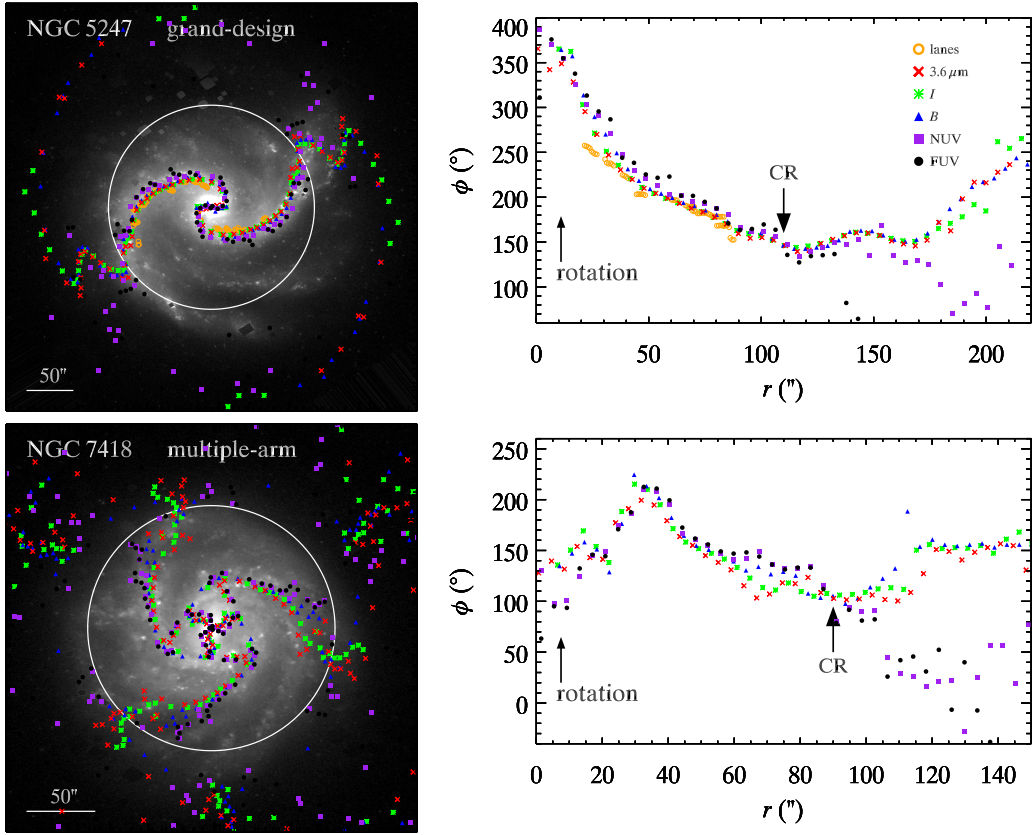}
\caption{Results of 1D Fourier decomposition for NGC~5247 (upper row) and NGC~7418 (bottom row). 
Left panels present the CGS {\it B}-band images, overplotted with the resulting centroids of the spiral arms, 
with their polar coordinate shown in the right panels, in $3.6\,\micron$ (red crosses), {\it I}\ (green stars), 
{\it B}\ (blue triangles), NUV (purple squares), and FUV (black points).  
Right panels show the phase angles of spiral arms, identified by the 1DDFT method, as a function of galactocentric radii. 
We adopt the convention that the azimuthal angle increases with rotation.
The orange symbols, if available, represent the dust lanes, identified visually in the {\it B}-band images. 
The white circles in the left panels indicate the estimated corotation radius (CR), at which location 
the color gradient reverses; the arrow in the bottom left corner of the right panels denotes the direction of rotation.
}
\end{figure}

\begin{figure}
\figurenum{2}
\centering
\includegraphics[width=14cm]{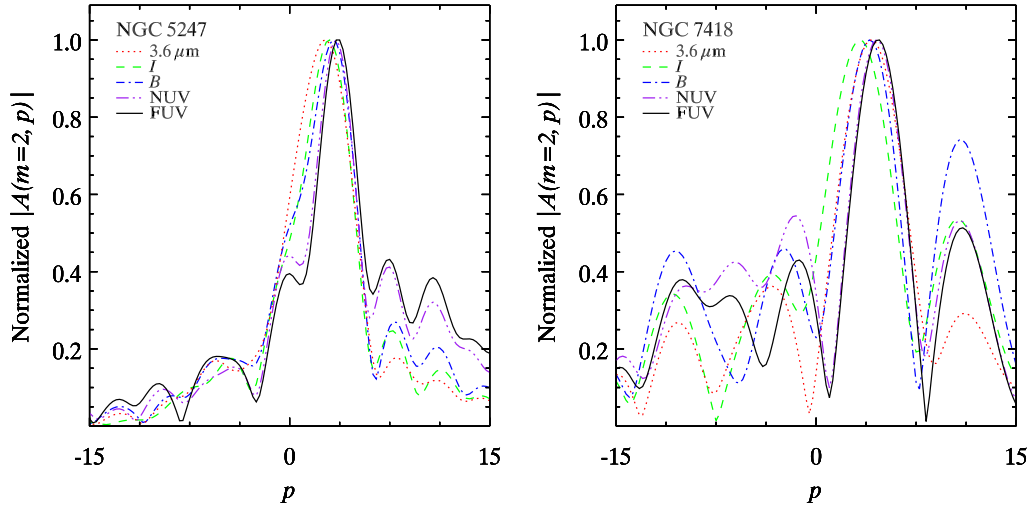}
\caption{2D Fourier spectra over the radial ranges 
[15$\arcsec$,201$\arcsec$] for NGC~5247 (left)
and [21$\arcsec$,126$\arcsec$] 
for NGC~7418 (right), 
set to be the same for all bandpasses. 
The spectra for $3.6\,\micron$, {\it I}, 
{\it B}, NUV, and FUV are marked by the red dotted, green dashed, blue dot-dashed, purple 
dot-dot-dashed, and black solid lines, respectively.  As the bandpass becomes bluer, the spectra 
tend to shift gradually to larger value, i.e., the pitch angle becomes smaller (see Eq. (3) and Section 4.1).
}
\end{figure}

\section{Pitch Angle Measurements}

The most widely used techniques to measure arm pitch angle employ discrete
Fourier transformation, either in 1D (1DDFT) \citep{Grosbol2004, Kendall2011}
or in 2D (2DDFT) \citep{Kalnajs1975, Iye1982, Krakow1982, Puerari1992,
Puerari1993, Block1999, Seigar2005, Davis2012}.  Both techniques are used and
discussed in detail by \cite{Yu2018} in the context of the CGS images.
Here we just briefly summarize a few essential points.

The 1DDFT method fits a Fourier series to the 1D azimuthal isophotal light
profile to identify the centroids of the spiral arms using the phase 
angle of the dominant Fourier mode $m$.  When computing the azimuthal isophotal
light profile (using the IRAF task {\it ellipse}), in addition to keeping $e$, 
PA, and galaxy center fixed to the reference values from the {\it I}-band 
image, it is also essential to adopt the same physical linear step size for 
the isophotes.  If the spiral arms are correctly identified by the phase angle 
of the Fourier mode, the phase angle profile as a function of radius will show 
almost monotonic change, whose gradient reflects the pitch angle of spiral 
arms.  Fitting a logarithmic function to the phase angle as a function of 
radius, $\phi_{m}(r)$,

\begin{eqnarray}
\phi = b \cdot {\rm ln}\, r + {\rm constant},
\end{eqnarray}

\noindent
where $r$ is the radial distance from the center and $b$ is a coefficient.  
The pitch angle follows, $\varphi$, from

\begin{eqnarray}
\varphi = \text{arctan}\left( \frac{1}{b} \right),
\end{eqnarray}

\noindent
with its error determined through propagation of the fitting error of $b$.
This procedure, of course, cannot be applied to totally flocculent, irregular or
non-symmetric spiral patterns, which lack a smooth phase angle profile.  Even
for objects that do possess a well-behaved phase angle profile, the radial
range over which it changes smoothly and monotonically will not be identical in
every band because of variations in the noise properties of the images.  This
is illustrated in Figure 1 for the grand-design spiral NGC~5247 and the
multiple-arm spiral NGC~7418, showing the phase angle profiles of the 
dominant Fourier mode calculated for images in $3.6\,\micron$,
{\it I}, {\it B}, NUV, and FUV ({\it R} and {\it V}\ bands are omitted for 
clarity). 
Note that we adopt the convention that the azimuthal angle increases
with rotation.
NGC~7418 has three inner arms associated with the central bar, with 
one of them disappearing in the outer part. 
Because its 1D Fourier components show highest amplitude in $m$\,=\,3 while
2D Fourier components show highest amplitude in $m$\,=\,2, 
a different Fourier mode 
is used in the 1DDFT ($m=3$) and 2DDFT ($m=2$; Figure 2) methods to calculate pitch angle; 
both give consistent results. 
Some multiple-armed galaxies ($\sim$\,10\% in our sample), which have more 
complicated spiral structures than grand-design galaxies, 
may exhibit a different dominant mode in 1D and 2D Fourier analysis.
In other words, this kind of galaxies has 
no ``true'' dominant mode.
The right panels show that the centroids of 
the spiral arms drift downstream gradually and systematically from $3.6\,
\micron$ to the FUV (this will be discussed in further in the next section).
The {\it GALEX} AIS FUV and NUV images have particularly poor S/N. 
To minimize potential uncertainty from variation of pitch angle with radius, we 
restrict the analysis in each band to the same radial range
over which all the phase angle profiles change almost monotonically.
This radial range may only occupy part of the full spiral region, so 
that the pitch angle measured by us is not necessarily consistent with the intrinsic
pitch angle of spiral arms for those galaxies whose pitch angle varies 
significantly with radius.
Using the 1D method, we find for NGC~5247 pitch angles of $33\fdg7\pm0\fdg8$, 
$35\fdg2\pm0\fdg6$, $31\fdg2\pm0\fdg7$, $31\fdg0\pm0\fdg7$, and 
$29\fdg3\pm1\fdg0$ in $3.6\,\micron$, {\it I}, {\it B}, NUV, and FUV, 
respectively; for NGC~7418, we measure $32\fdg8\pm2\fdg7$, 
$31\fdg4\pm1\fdg8$, $30\fdg4\pm1\fdg4$, $30\fdg6\pm3\fdg0$, and 
$28\fdg8\pm2\fdg6$ in $3.6\,\micron$, {\it I}, {\it B}, NUV, and FUV, 
respectively. 
Using the 1D method, we successfully measure
pitch angles for 71 3.6\,$\mu m$, 60 NUV, and 50 FUV images of 96 galaxies (Table 1).
Typical uncertainties in the measurements are $\sim 2\degr$.

The 2DDFT method decomposes the background-subtracted, star-cleaned, and 
deprojected image, transformed into polar coordinates, into a superposition of 
2D Fourier components 
with coefficients $|A(m,p)|$
\footnote{$A(m, p) = \frac{1}{D} \int ^{\text{ln}(r_{\text{out}})}_{\text{ln}(r_{\text{in}})} \int^{\pi}_{-\pi} \sum_{j=1}^{N} I_j(r_j, \theta_j) \delta(\mu-\mu_j) \delta(\theta-\theta_j) e^{-i(m\theta + p \mu)} d\theta d\mu,$
where $D$\,=\,$\sum_{j=1}^N$$I_j$, 
$I_j$ is intensity of the $j$th pixel at $(r_j, \theta_j)$, 
$\mu \equiv {\rm ln}$\,$r$, 
$m$ is the azimuthal frequency, and $p$ is the logarithmic radial frequency.
}
calculated over the four radial ranges determined 
by \cite{Yu2018} in the {\it I}\ band: 
[$r_{\rm in}$, $r_{\rm out}$], [$r_{\rm in}$+0.2$\Delta r$, $r_{\rm out}$], 
[$r_{\rm in}$, $r_{\rm out}$-0.2$\Delta r$], and [$r_{\rm in}$+0.1$\Delta r$, $r_{\rm out}$-0.1$\Delta r$], 
where $\Delta r$\,=\,$r_{\rm in}$\,$-$\,$r_{\rm out}$, 
and $r_{\rm in}$ and $r_{\rm out}$
are the inner and outer boundaries of the spiral arms (Table 1), which are set the same for all seven bandpass images. 
The peak $p^{\prime}$ of the power spectrum of 
the dominant Fourier mode $m$, which is set the same for images in 
all seven bands, is identified to calculate the arm pitch angle:

\begin{eqnarray}
\varphi = \arctan{\left(- \frac{m}{p^{\prime}} \right)}.
\end{eqnarray}

\noindent
The pitch angle of the galaxy is taken as the mean value of the pitch 
angle in the four radial bins; its uncertainty, typically $\sim 3\degr$, is 
their standard deviation. 
The Fourier spectra, $|A(m$\,=\,$2,p)|$, for NGC~5247 and NGC~7418 are
illustrated in the left and right panels of Figure 2, respectively. 
The $p^{\prime}$ corresponding to the most prominent peak gradually shifts 
toward the right with wavelength, indicating smaller pitch angles. For 
NGC 7418, although the peak at 3\,$\mu$m is not in the canonical order, 
there is an important shift toward larger values of $p^{\prime}$, especially between the {\it I} and FUV bands. 
The 2D method yields for NGC~5247 pitch 
angle $36\fdg9\pm6\fdg5$, $37\fdg9\pm5\fdg7$, $35\fdg9\pm7\fdg2$, 
$30\fdg7\pm3\fdg9$, and $28\fdg5\pm3\fdg6$ in $3.6\,\micron$, {\it I}, {\it B},
NUV, and FUV, respectively; the corresponding values for NGC~7418 are
$32\fdg7\pm3\fdg7$, $32\fdg3\pm3\fdg7$, $25\fdg3\pm3\fdg1$, $27\fdg3\pm6\fdg7$,
and $22\fdg6\pm3\fdg7$. 
Using the 2D method, we successfully measure
pitch angles for 69 3.6\,$\mu m$, 70 NUV, and 60 FUV images of 99 galaxies (Table 1).

\begin{figure}
\figurenum{3}
\centering
\includegraphics[width=17.5cm]{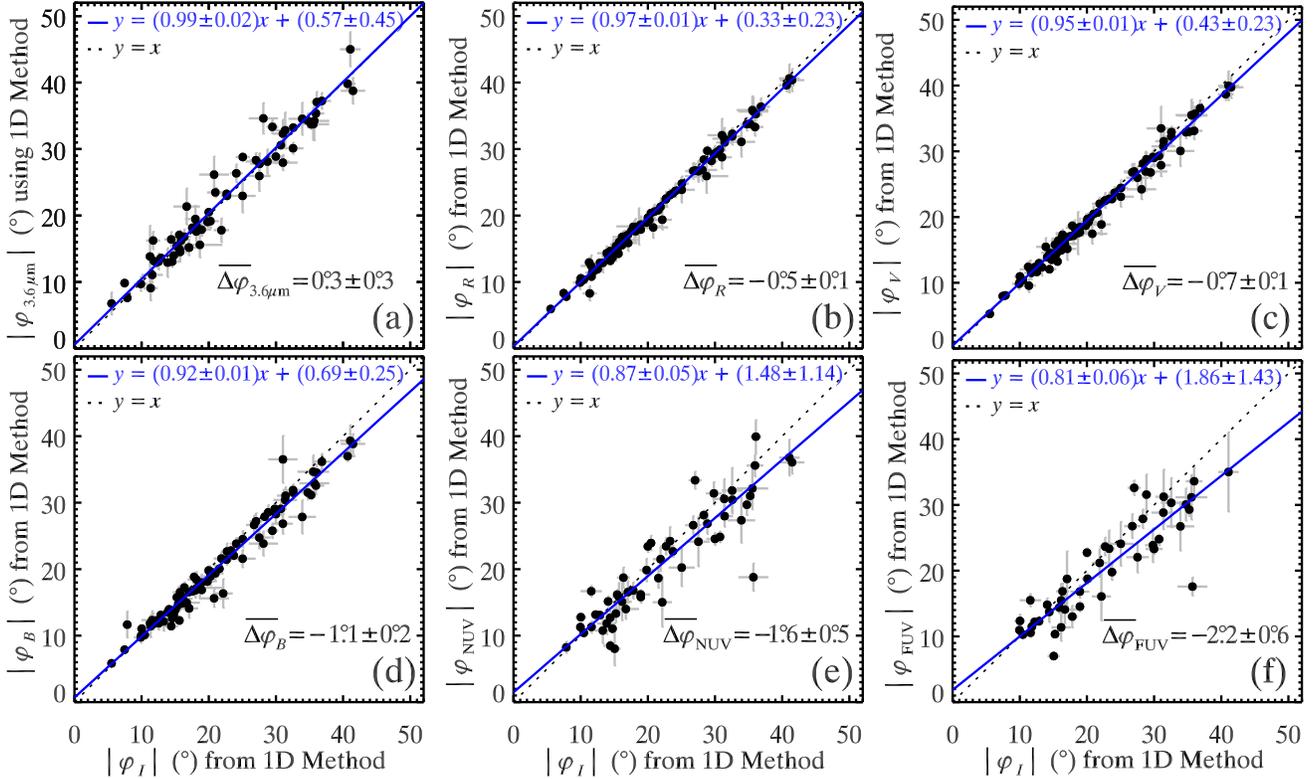}
\caption{Correlations between the pitch angle of spiral arms, calculated from the 1D method, in the {\it I}\ band and in the (a) 3.6 $\micron$, (b) {\it R},  (c) {\it V}, (d) {\it B}, (e) NUV,  and (f) FUV bands.  In each panel, the dashed black line marks the 1:1 relation, and the blue solid line is the best-fit straight line. The mean difference, $\overline{\Delta\varphi}$,  and its error are given in the bottom-right corner of each panel.}
\end{figure}

\begin{figure} 
\figurenum{4}
\centering
\includegraphics[width=17.5cm]{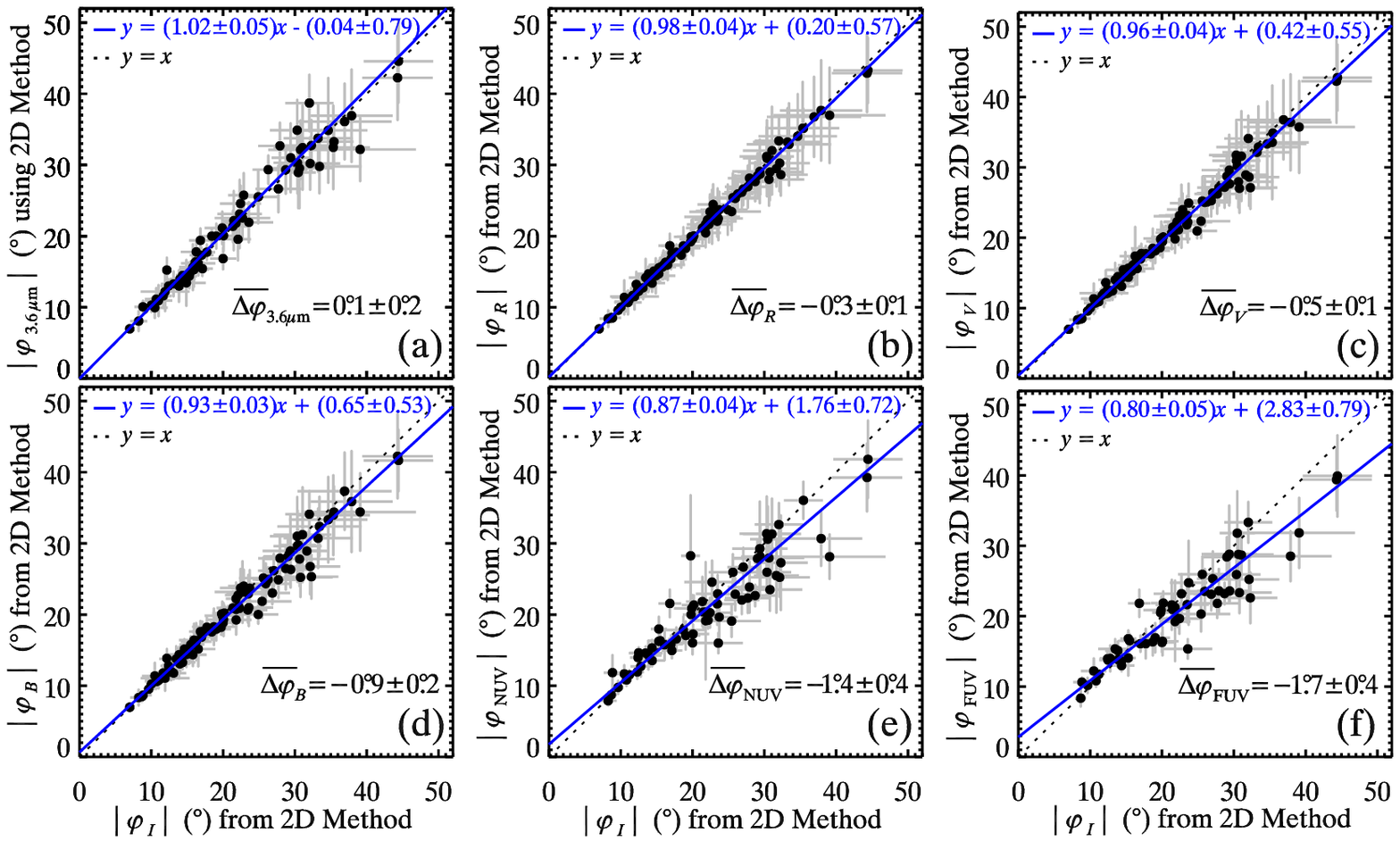}
\caption{All symbols as in Figure 3, but for pitch angles calculated from the 2D method.}
\end{figure}

\section{Results}
\subsection{Color gradient across spiral arms}

Young stars triggered by the spiral shock will form an SF-arm, which is downstream from
the P-arm comprised of evolved stars, resulting in a color jump from
red to blue in the direction of rotation inside corotation radius \citep{Gittins2004}.
The classic age gradient, from blue to red, occurs as these newly formed young stars drift differentially out of the
SF-arm and age meanwhile. Both kinds of color gradients are caused by
density waves and can coexist. However, the techniques we have applied 
(1D and 2D Fourier transforms) implicitly use the intensity as weighting when calculating Fourier
components. Thus, these methods tend to identify the centroid
of spiral arms close to their brightness peak and are not sensitive to the age color 
gradient, 
which occurs in a wider azimuthal scale as the surface brightness is gradually diminishing. 
As shown in Figure 1, we witness the centroid of spiral 
arms gradually drifting downstream in waveband from $3.6\,\micron$ to FUV, 
resulting in a continuous color gradient from red to blue in the direction of 
rotation inside the corotation radius. This is consistent with the prediction by \cite{Gittins2004}, 
although the observed gradient is not obviously discontinuous.  Such a color gradient 
has also been detected by \cite{Martinez2013} using the phase angle (not 
azimuthal position) of the dominant mode from 2D Fourier transformation, even 
though the authors have treated it mistakenly as a result of an age gradient.
The detected continuity of the color gradient 
is probably a consequence of dust extinction. 
The spiral shock is located upstream from the SF-arms and can be traced by
the dust lanes \citep{Gittins2004}. 
The light from young stellar complexes on the 
side closer to the shock is delimited by extinction 
(some of the young stellar complexes are actually still visible at 3.6\,$\micron$).
Therefore, if the dust lanes are tapering downstream from the shock inside
corotation radius, at progressively shorter wavelengths, arms will be increasingly 
located further from the shock, since they will be visible 
only once the optical depth $\tau_{\lambda}$ falls below 1.
The stars and gas are still moving relative to the spiral pattern as expected from
density wave theory, so the width of the arm has to vanish at corotation and the pitch angle
has to be smaller for the bands detected further away from the shock.

We identify the dust lanes, which are marked by orange symbols in Figure 1, 
from visual inspection of the {\it B}-band image of NGC~5247, while there is no 
regular dust lane for NGC~7418. The dust lanes are not as 
regular and continuous as 
the shock fronts generated in numerical simulations employing a 
rigidly rotating spiral potential \citep[e.g.,][]{Gittins2004, KimKim2014},
and they only occupy part of the full spiral region. Despite these caveats, 
some weak evidence of dust lanes being upstream from the spiral arms is found.
The Fourier spectra presented in Figure 2 also shed light on this picture. 
For NGC~5247, the $|A(m$\,=\,$2,p)|$ peak profiles get narrower with decreasing wavelength.
For NGC~7418, whose dust lanes are inconspicuous, the $|A(m$\,=\,$2,p)|$ peaks are 
equally wide and quite symmetric at all wavelengths.
We highlight that, with or without dust extinction, the detected color gradient 
as well as the trend that spiral arms are more tightly wound in progressively shorter wavelengths (see Section 4.2) 
are consistent with spiral density waves.
Outside of the corotation radius, the stars and gas clouds move slower than the spiral 
pattern, leading to a reversed color gradient. Based on this behavior, the 
coratation radius can be estimated as the radius at which the color gradient 
reverses; it is marked with the white circle in the left panels and with the label CR in the 
right panels of Figure~1.

\subsection{Wavelength dependence of pitch angle}

Figure 3 shows the pitch angles of spiral arms measured using the 1D method in 
the $I$ band, compared with pitch angles derived in the 3.6 $\micron$, {\it R},
{\it V}, {\it B}, NUV, and FUV bands.  The corresponding results obtained from 
the 2D method, which are entirely consistent with those from the 1D method, 
are shown in Figure 4.  In each panel the dashed line represents the 1:1 
relation, and the best-fit straight line, whose functional
form is given at the top, is marked by the blue solid line.  We calculate the 
mean difference of pitch angle between waveband $\lambda$ and the 
{\it I}\ band, $\overline{ \Delta \varphi_{\lambda}} = 
\sum{(\Delta\varphi_{\lambda, i})}/N = \sum{(|\varphi_{\lambda, i}|-
|\varphi_{{\it I}, i}|)}/N$, where $\lambda$ can be 3.6 $\micron$, {\it R}, 
{\it V}, {\it B}, FUV, or NUV, $i$ denotes the $i$th galaxy, and $N
$ is the total number of galaxies. The error of the mean is 
calculated by $\epsilon = (\text{standard deviation of}\, 
\Delta\varphi_{\lambda,i})/\sqrt{N}$.

Both the 3.6 $\micron$ and $I$-band images give essentially identical 
pitch angles for the spiral arms, for both the 1D (Figure 3a) and 2D (Figure 4a) 
methods.  The best-fit straight line 
for these two wavelengths has a slope of $0.99$ 
($1.02$) and an intercept of $0\fdg6$ ($0\fdg0$); the mean difference is 
$\overline{\Delta \varphi}_{3.6\mu m} = 0\fdg3 \pm 0\fdg3$ ($0\fdg1 \pm 0\fdg2$).  The 
excellent consistency between these two long-wavelength bandpasses confirms 
that the $I$ band traces the mass distribution of evolved stars just as 
effectively as the 3.6 $\micron$ band.  This is a crucial step, for it allows 
us to use the CGS $I$-band images as the reference red bandpass with which to 
compare the other five bluer bandpasses.  The $I$-band images have the 
advantage of having much higher signal-to-noise ratio and angular resolution 
than the 3.6 $\micron$ IRAC images.  {\it I}-band images closely match 
the conditions of the {\it BVR}\ images, and there are also many more objects 
that overlap with the {\it GALEX} database. We detect a strong 
correlation between the $I$-band pitch angle and the pitch angle measured in 
each of the bluer bands, with a slope for the best-fit straight line 
that decreases consistently and systematically, from $0.97$ ($0.98$) in the 
$R$ band to $0.81$ ($0.80$) in the FUV band (Figures 3 and 4).
The correlations are tightest for the optical bands because of the uniformly 
high quality of the CGS data, while the relatively low signal-to-noise ratio 
and coarse pixel scale of the FUV/NUV images from {\it GALEX}/AIS introduce 
greater scatter but negligible systematic bias (see Section 4 of \citealt{Yu2018}).  
Meanwhile, the mean difference in pitch angle with respect to the $I$ 
band decreases (the absolute value of difference increases) systematically and significantly 
(at greater than the 3 $\sigma$ level) from $\overline{\Delta\varphi}_{\it R} = 
-0\fdg5\pm0\fdg1$ ($-0\fdg3\pm0\fdg1$) in the {\it R} band to 
$\overline{\Delta\varphi}_{\rm FUV} = -2\fdg2\pm0\fdg6$ ($-1\fdg7\pm0\fdg4$) in the FUV 
band.  This clear statistical trend implies that spiral arms in bluer 
bandpasses are, on average, tighter than in redder bandpasses.  

The absolute value of the pitch angle difference is very small.  We can 
discern it only when the inter-band correlations have a sufficiently small 
total scatter, which we achieved through adopting a consistent approach to 
measuring pitch angles across different bands (Section 3).  This may explain 
why \cite{Davis2012} failed to detect a discernible difference in pitch angles 
between the {\it I}\ and {\it B} bands, even though they also analyzed CGS 
data. Our results agree with those of \cite{Grosbol1998}, \cite{Martinez2012}, 
\cite{Martinez2013}, and \cite{Martinez2014}, but we place them on a firmer 
statistical footing using a much larger and more comprehensive sample.  
By contrast, \cite{PI2016} reported an entirely opposite result: in their 
analysis of spiral arms of 28 galaxies, they found that the pitch angles are 
actually {\it larger}\ in the FUV band than at $3.6\,\micron$.  Moreover, the 
mean difference in pitch angle they measured is huge: $\overline{\Delta\varphi} = 13\degr$ 
(recalculated from data in their table).  Their pitch angle measurements, 
unfortunately, are not reliable. As an example, for the well-studied 
grand-design spiral galaxy NGC 1566, which overlaps with our sample, 
\cite{PI2016} quote a pitch angle $\varphi = 15\fdg29$ for 3.6 $\micron$, 
$31\fdg20$ for $B$, and $44\fdg13$ for FUV.  However, simple 
inspection of the images of NGC 1566 clearly reveal that its two symmetric
arms are quite similar in $3.6\, \micron$, {\it B}, and FUV. The 
huge difference of pitch angle reported by \cite{PI2016} for this galaxy cannot 
possibly be correct.  Inspection of other galaxies in their sample (e.g., 
NGC 1097) reveals similar problems.  As mentioned earlier, it is crucially 
important to adopt consistent parameters when measuring pitch angles across 
different images, especially those having vastly different angular resolution 
and signal-to-noise ratio.  We suspect that these factors may have adversely 
affected the measurements of \cite{PI2016}.

Our finding that the pitch angle of spiral arms decreases toward shorter 
wavelengths implies that young stars follow tighter arms than those delineating
the old stellar population. 
This behavior arises naturally as a consequence of star formation triggered by the 
large-scale galactic shocks of trailing spirals and the fact that our techniques 
are intensity-weighted. 
The oldest stars, as traced 
in the $I$ or $3.6\,\micron$ band, form a long-lived spiral pattern with a 
unique pattern speed and generate the spiral gravitational potential (P-arm). 
The gravitational potential perturbs the gas orbits, produces spiral shocks in 
the gas flow, and triggers gravitational collapse of the gas to form new stars 
to enhance star formation (SF-arm) downstream from the P-arm for 
trailing spirals inside corotation radius \citep{Gittins2004}.  
These newly formed stars drift out of the arm and age meanwhile. 
After aging $\sim10^7{\rm years}$, the stars will move further downstream from the SF-arm. 
For trailing spirals, within the corotation radius and in the direction of 
rotation, there should a spatial ordering across the spiral arm: spiral shock, 
the oldest stars, young stars, and aging stars. 
However, because the techniques we employ (1D and 
2D Fourier transformation) implicitly use the intensity as weighting when 
calculating Fourier components, they tend to identify the centroid 
of spiral arms close to its peak to measure the pitch angle of the SF-arms and P-arms, 
and are thus not sensitive to the age color gradient. 
We therefore detect a color gradient from red to blue (Figure 1) and 
tighter arms in bluer bands (Figures 3 and 4).  Dust extinction causes the 
color gradient and the variation of pitch angle with bandpass to appear 
continuous, instead of discontinuous as predicted by \cite{Gittins2004}.

Several factors complicate this simple picture.  Recent works using a 
generalization of the Tremaine-Weinberg method \citep{TW1984} to calculate the 
pattern speed propose that the spiral pattern speed may increase with 
decreasing radius in some objects \citep{Merrifield2006, Meidt2009, 
Speights2012}. This would reduce the expected offsets between the tracers of 
the different stellar populations 
and imply that, actually, there is no pattern speed. 
However, an increase in the measured pattern 
speed with decreasing radius might be caused by the larger inward streaming 
motions of young stars at smaller radius rather than by a real radial variation of 
$\Omega_{\rm P}$ \citep{Martinez2009b}. The position of the spiral
shock may depend on the pattern speed, strength of the spiral potential, and 
the detailed properties of the gas clouds \citep{Gittins2004, KimKim2014}.  Finally, 
galaxies possess a variety of rotation curves, from slow-rising shapes for 
late-type galaxies to steeply rising and then flat shapes for early-type 
ones \citep[e.g.,][]{Kalinova2017}.  Thus, for any realistic sample of 
galaxies, such as that considered in this study, all of these factors will 
introduce scatter to the predicted simple variation of spiral pitch angle with 
stellar population. 
Yet, remarkably, we still seem to be able to extract a clear signal consistent with the predictions. 
It is possible that in this case, among all the effects diluting the signal, extinction is actually helping it.

Although our results indicate that spiral arm structure in our sample is 
long-lived as predicted by density wave theory, it is still possible that some 
galaxies are dominated by transient but recurrent spiral structure obeying 
swing amplification theory \citep{Toomre1981}. This scenario, which does not 
anticipate color gradients across spiral arms, may account for the objects 
located closest to the 1:1 line in Figures 3 and 4.

\begin{figure}
\figurenum{5}
\centering
\includegraphics[width=17.5cm]{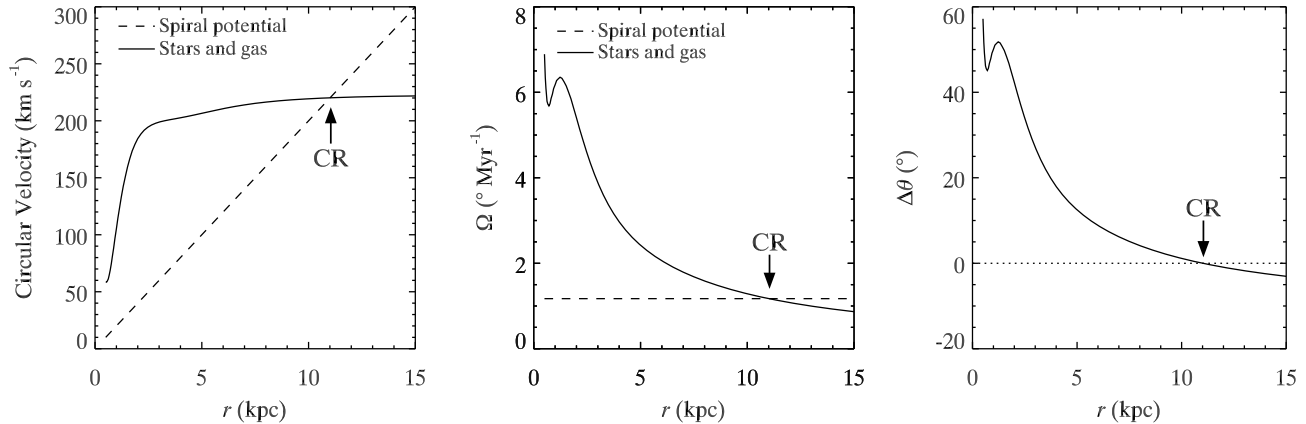}
\caption{Simplified model to explain the observed angular offset between the 
SF-arm and the P-arm. Left: rotation curve of the stars or gas (solid line), 
and spiral pattern with pattern speed $\Omega_{\rm P}=20$~km\,s$^{-1}$\,kpc$^{-1}$ 
(dashed line). Middle: angular velocity as a function of radius for 
spiral potential and for stars or gas. Right: the azimuthal offset 
between the arms of newly formed young stars (SF-arm) and the spiral 
potential (P-arm), assuming a star formation timescale of 
$\uptau$ = 10 Myr: $\Delta\theta=\theta_{\rm SF}-\theta_{\rm P}=
(\Omega_{\rm stars/gas}-\Omega_{\rm P})\times\uptau$.  The dotted line indicates 
zero offset.}
\end{figure}

\begin{figure}
\figurenum{6}
\centering
\includegraphics[width=17.5cm]{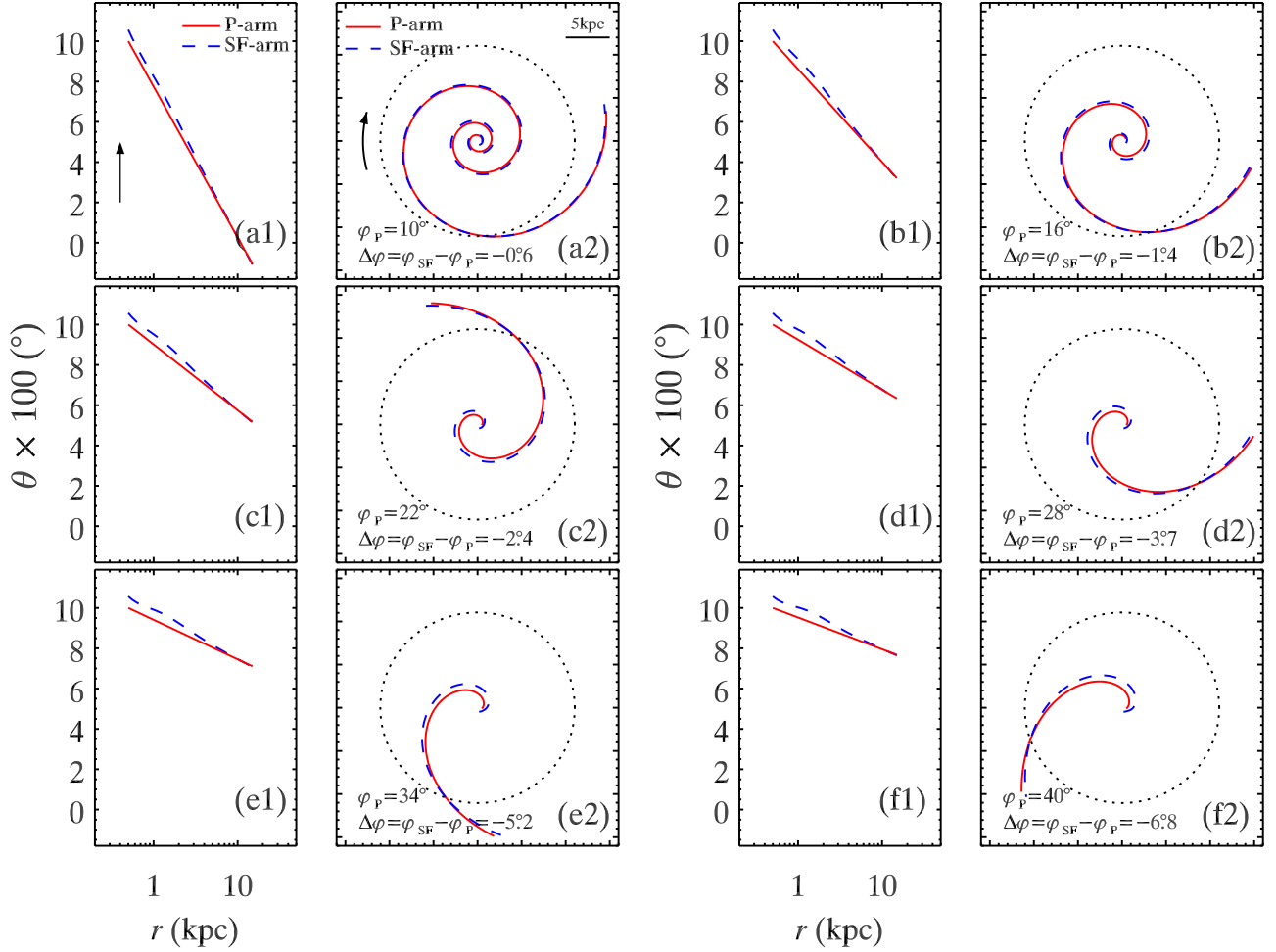}
\caption{Predicted location of the P-arm and S-arm, based on the model 
assumptions given in Figure 5.  The solid red line represents the spiral potential (P-arm), with 
the pitch angle shown in the legend, while the dashed blue line represents the 
arm comprised of newly-born massive stars (SF-arm), triggered by the 
spiral shock, with an estimated timescale of 10 Myr. Panels (a1)--(f1) present 
the spiral arms in (ln\,$r$, $\theta$) space, while panels (a2)--(f2) present 
them in Cartesian coordinates, with a dotted circle indicating the 
corotation radius. The difference of pitch angle between the SF-arm 
and P-arm, $\Delta\varphi=\varphi_{\rm SF}-\varphi_{\rm P}$, is given in the 
bottom-left corner of panels (a2)--(f2). The arrows indicate the direction of 
rotation.}
\end{figure}

\begin{figure}
\figurenum{7}
\centering
\includegraphics[width=13.5cm]{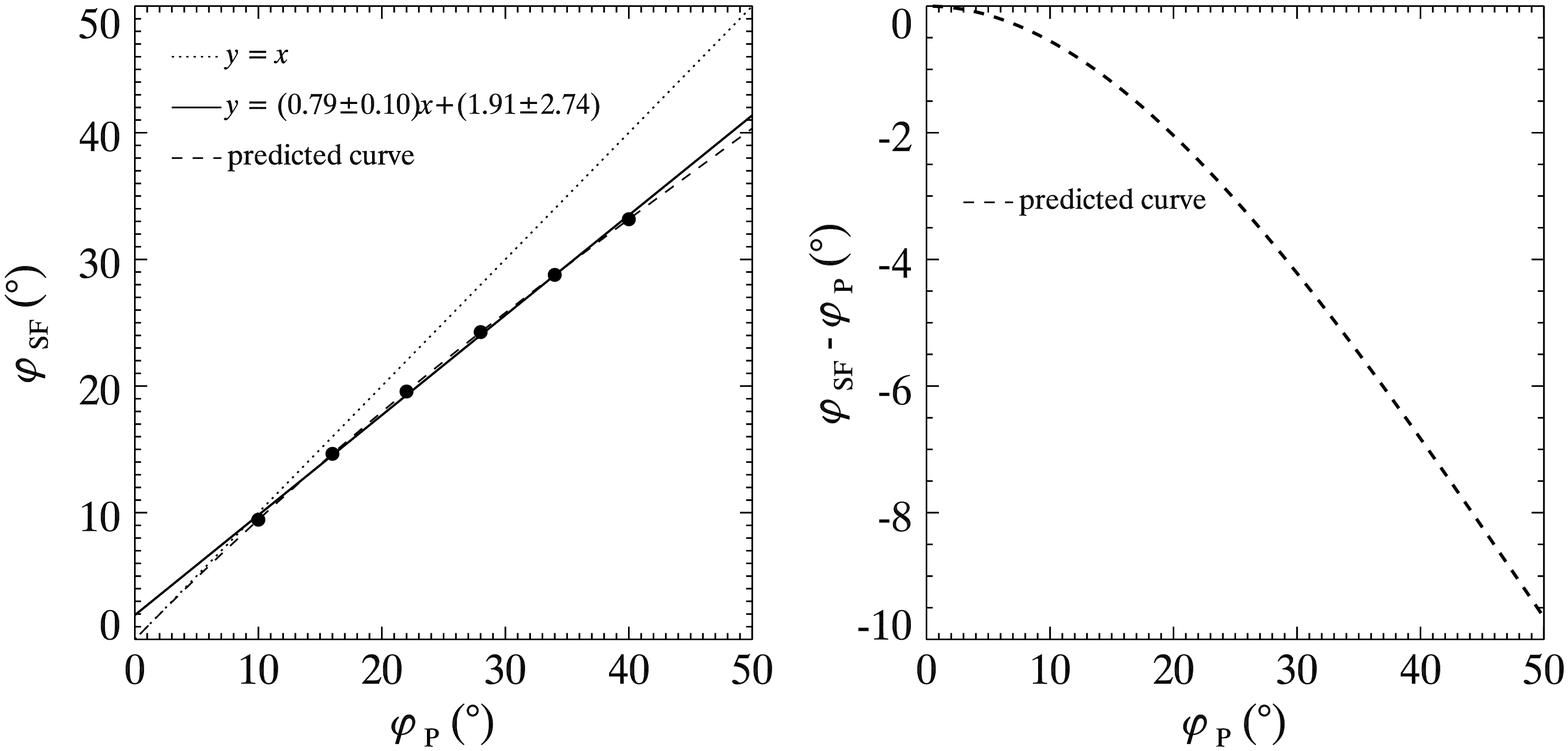}
\caption{Left: relation between SF-arm pitch angle ($\varphi_{\rm SF}$) 
and P-arm pitch angle ($\varphi_{\rm P}$). Right: difference  
between the two pitch angles, $\varphi_{\rm SF}-\varphi_{\rm P}$, as a function of $\varphi_{\rm P}$.
The behavior predicted by the model shown in Figure 5 is plotted as 
the dashed curve; the data points are the results from Figure 6, whose best-fit 
function is plotted as the solid line.  The dotted line represents $y=x$.}
\end{figure}

The correlation between $I$-band pitch angle and FUV-band pitch angle (Figures 
3f and 4f) has a best-fit slope of $\sim 0.8$, which implies that the absolute 
value of the difference of pitch angle between young and old stars increases 
progressively with increasing pitch angle.  Moreover, the intercept is greater 
than zero.  Both of these characteristics need explanation.

\subsection{Models}
We construct a simplified model to explain these observations.  The 
pitch angle of spiral arms on average increases with later Hubble type, 
but with a large variation in pitch angle for any given Hubble type 
\citep{Kennicutt1981, Ma2002, Yu2018}. 
As pitch angle shows no conclusively clear, strong correlation with global galaxy properties
\citep{Kennicutt1981, Seigar2005, Kendall2015, Yu2018}, we construct a 
simplified model by choosing a typical value of pattern speed, rotation curve, 
and star formation timescale.  The compressed gas is traditionally believed to 
be located upstream from the potential inside corotation radius, but simulations
show that its location relative to the potential is complex and may depend on 
the sound speed of the gas, the strength of the spirals, or the pattern 
speed.  The compressed gas can be downstream from the potential if it is cold or multiphase 
\citep{Gittins2004, Wada2008}, or if the strength of the spirals is weak 
\citep{KimKim2014}.  As we do not find a significant offset in azimuthal angle 
between the dust lanes and the centroid of the arms at $3.6\,\micron$ (e.g., 
Figure 1), we assume that the location of the spiral shock is very near to 
the minimum of the spiral potential.

Our model adopts a spiral potential (P-arm), comprised of the oldest 
stars, with pattern speed $\Omega_{\rm P}=20$\,km\,s$^{-1}$\,kpc$^{-1}$.  
This value is consistent with that found by \cite{Martos2004} for their optimal
dynamically self-consistent model of the Milky Way, and it lies within the 
range of pattern speeds of spiral arms (10 to 45\,km\,s$^{-1}$\,kpc$^{-1}$) 
measured by using the Tremaine-Weinberg method \citep[e.g.,][]{Zimmer2004, 
Fathi2009} or by analyzing age color gradients \citep{Martinez2011}.
The stars and gas in the disk are assumed to rotate following the circular 
velocity profile shown in Figure 5 (left), which is constructed using the 
eigenstates derived from the principle component analysis of the rotation 
curves in the CALIFA galaxy sample \citep{Kalinova2017}. The circular velocity 
profile, like that of the Milky Way \citep[e.g.,][]{Bovy2012}, rises rapidly 
to $\sim$\,$200$\,km\,s$^{-1}$ by $r$\,$\approx$\,3 kpc and then flattens to $\sim$\,220 
km~s$^{-1}$ beyond $r$\,$\approx$\,10\,kpc.  The angular velocity profile 
is shown in the middle panel of Figure 5. As young, massive stars form on a 
short timescale of \lax\ 10 Myr, we assume that this is roughly also the 
timescale ($\uptau$) for gas clouds to get shocked, form massive stars, 
and appear as star-forming arms (SF-arm) observed in the FUV images.
We will compare results from the model with 
our results for $\overline{\Delta\varphi}_{\rm FUV}$, the pitch angle offset between the 
{\it I}-band and FUV images.
Although the resolution of the FUV images is coarser, the spiral arms are still well-resolved 
because these galaxies are very nearby.
As illustrated in Figure 5 (right), the SF-arm is downstream from the 
P-arm inside the corotation radius and has an azimuthal angle offset relative 
to the P-arm $\Delta\theta=\theta_{\rm SF}-\theta_{\rm P}=(\Omega_{\rm star/gas}
-\Omega_{\rm P})\times\uptau$.  Large negative angular offsets are expected in 
the inner regions, and the offsets become positive beyond the corotation radius.

To mimic the properties of our sample, whose spiral pitch angles range from 
$10\degr$ to $40\degr$, we homogeneously sample the pitch angle the of 
P-arm at six equally spaced intervals of 10$\degr$, 16$\degr$, 22$\degr$, 
28$\degr$, 34$\degr$, and 40$\degr$ (Figure 6).  The P-arm is assumed 
to be logarithmic in shape, following the equation:
\begin{eqnarray}
\theta_{\rm P} = b\,{\rm ln}\,r_{\rm P} + a,
\end{eqnarray}
\noindent
where ($r_{\rm P}$, $\theta_{\rm P}$) are the polar coordinates of the 
minima of the spiral potential, 
$a$ is a quantity determining the phase angle of the arm, and 
$b$ is a parameter defining pitch angle, $\varphi={\rm tan}^{-1}(\frac{1}{b})$.
Considering only circular motion, $r_{\rm SF}=r_{\rm P}$ and 
$\theta_{\rm SF}=\theta_{\rm P}+\Delta\theta$. 
As the resulting SF-arm, traced by the blue dashed curve in Figure 6, slightly 
deviates from logarithmic shape, a function with the mathematical form of 
Eq. (4) is fitted to the SF-arm to obtain the arm pitch angle, and then the
offset in pitch angle between the SF-arm and P-arm is calculated as
$\Delta \varphi=\varphi_{\rm SF}-\varphi_{\rm P}$.  This is presented in each 
of the panels of Figure 6. Consistent with our observations, the SF-arm has 
smaller pitch angle---and hence is tighter---than the P-arm.

We further fit a straight line to the six data points in the left panel of Figure 7 to find the best-fit function, 
marked by the solid line. 
The predicted relation between $\varphi_{\rm SF}$ and $\varphi_{\rm P}$
from the model, shown as the dashed line, is a non-linear function, but it agrees
extremely well with the fit between {\it I}-band and FUV pitch angle. The slope (0.79) and intercept (1.91) of the 
best-fit theoretical relation are entirely consistent with our observations: slope 
$0.81\pm0.06$ and intercept $1.86\pm1.43$ for the 1D method (Figure 3f), slope 
$0.80\pm0.05$ and intercept $2.83\pm0.79$ from the 2D method (Figure 4f).  
The positive intercept of the best-fit lines (Figures 3, 4, and 7) is 
an artifact caused by fitting a straight line to the data;
it is unphysical because it would prevent the arm from vanishing at corotation. 
As our sample has a deficit of pitch angles less than 10$\degr$, this behavior needs 
to be verified with further observations.
The slopes $<$ 1, on the other hand, reflect the fact that the absolute value of 
$\varphi_{\rm SF} - \varphi_{\rm P} = \Delta \varphi$ increases with $\varphi_{\rm P}$ 
($\Delta \varphi$ becomes more negative). This correlation is shown in the right panel 
of Figure 7. In our model, the physics is the same for arms with different $\varphi$. 
Thus, the angular offsets between the P and SF arms, which depend on the rotation 
curve, pattern speed, and star formation timescale (i.e., on the physics), are the 
same for arms with different $\varphi$. The tendency of the absolute value of $\Delta \varphi$ 
to increase with $\varphi_{\rm P}$ is therefore a geometric effect. With increasing $\varphi$, 
for a fixed CR radius, the arms are shorter (Figure 6). Since the width of the arm must 
vanish at CR, the more open the P-arm is, the larger the reduction of $\varphi_{\rm SF}$ 
must be for this to happen in the available length.

\section{Summary and Conclusions}

In the context of density wave theory, star formation triggered by 
long-lived spiral arms generates a star-forming arm (SF-arm)
downstream from the potential arm (P-arm) of the oldest stars, 
and thus induce a color jump from red to blue in the direction 
of rotation inside corotation radius \citep{Gittins2004}. The burst of star formation   
enhanced by the spiral shock will age and drift further downstream
to form an opposite but coexisting color gradient (classic age gradient), from blue to red.
As both 1D and 2D Fourier transformation implicitly use intensity
as the weighting when calculating Fourier components, these techniques identify
the centroid of spiral arms, and the resulting pitch angle measurements gauge the tightness
of the SF-arm in blue bandpasses and of the P-arm in red bandpasses.
In other words, Fourier techniques will readily identify the 
Gittins \& Clarke jump,  
where bluer SF-arms are tighter, and be quite insensitive to the classic age 
gradient, which occurs on a much wider azimuthal scale, as the surface 
brightness continuously diminishes.
Evidence for the classic age color gradient has been found
by \cite{Gonzalez1996}, \cite{Martinez2009a}, and \cite{Martinez2011},
while contradictory results have also been reported \citep{Schweizer1976, 
Talbot1979, Foyle2011}.  
\cite{PI2016} looked for and claimed to have found the pitch angle trend expected from aging arms (i.e., smaller
pitch angle at redder wavelengths), but it is surprising they were successful 
while also using Fourier techniques.
Many attempts have also been made to detect the dependence of arm 
pitch angle on wavelength in the context of color gradient of \cite{Gittins2004}:
\cite{Grosbol1998} saw it, 
and explained it as caused by relative displacements of dust and young stars; 
\cite{Foyle2011} failed to detect the azimuthal offset between P-arms
and SF-arms based on a cross-correlation function analysis; 
\cite{Martinez2012} and \cite{Martinez2013} detected it but attributed their result
wrongly to aging; \cite{Martinez2014} reported a positive result for three out of five objects.

We perform a comprehensive analysis of this problem using
the largest sample to date of bright, nearby spiral galaxies from the 
Carnegie-Irvine Galaxy Survey (CGS), having high-quality optical ({\it BVRI}) 
images, combined with 3.6 $\micron$ {\it Spitzer}\ images that trace the 
underlying stellar mass distribution, and NUV/FUV {\it GALEX}\ images that 
probe sites of recent star formation.  We measure pitch angles using both 1D 
and 2D Fourier techniques developed and tested by \cite{Yu2018}.  As the 
expected signal is subtle, great care was taken to ensure that the 
measurements were made consistently across the images from different bands and 
instruments.

By using the centroid of spiral arms identified by Fourier transformation,
we show that the color across spiral arms of both grand-design and 
multiple-arm galaxies changes from red to blue in the direction of rotation 
inside the corotation radius.  
We demonstrate that there is almost no difference between the 
pitch angle of spiral arms as measured in the $3.6\,\micron$ band or in the 
$I$ band, both of which effectively trace the evolved stellar population in our
sample.  This important demonstration allows us to use the much larger sample 
of higher signal-to-noise ratio, higher resolution CGS $I$-band images as the 
reference with which all other bluer bandpasses can be compared.  We find that 
spiral arms are on average tighter (pitch angle $\varphi$ smaller) in the bluer
bandpasses than in {\it I}-band, at greater than 3 $\sigma$ significance, and the 
absolute value of difference 
increases systematically with increasing wavelength separation.  Referenced to 
the $I$ band, the mean difference in pitch angle, as measured from the 1D 
Fourier method, is $\overline{\Delta\varphi} = -0\fdg5\pm0\fdg1$ in $R$, 
$-0\fdg7\pm0\fdg1$ in $V$, $-1\fdg1\pm0\fdg2$ in $B$, $-1\fdg6\pm0\fdg5$ in 
NUV, and $-2\fdg2\pm0\fdg6$ in FUV.   The 2D method gives consistent results, 
with corresponding values of $\overline{\Delta\varphi} = -0\fdg3\pm0\fdg1$, 
$-0\fdg5\pm0\fdg1$, $-0\fdg9\pm0\fdg2$, $-1\fdg4\pm0\fdg4$, and 
$-1\fdg7\pm0\fdg4$.  
That $\overline{\Delta\varphi} < 0$ and the tendency for 
$\overline{\Delta\varphi}$ to become more negative with increasing $\varphi$ 
are both consequences of the fact the azimuthal angle offset between the 
spiral potential and the newly formed stars, triggered by the spiral shock, 
decreases with radius and vanishes at the corotation radius. Our results
support the density wave theory for the origin of symmetric spiral arms.

\begin{acknowledgements}
We thank the referee for constructive criticisms that helped to improve the 
quality and presentation of the paper.  This work was supported by the 
National Key R\&D Program of China (2016YFA0400702) and the National Science 
Foundation of China (11473002, 11721303). 
\end{acknowledgements}

\clearpage
\includegraphics{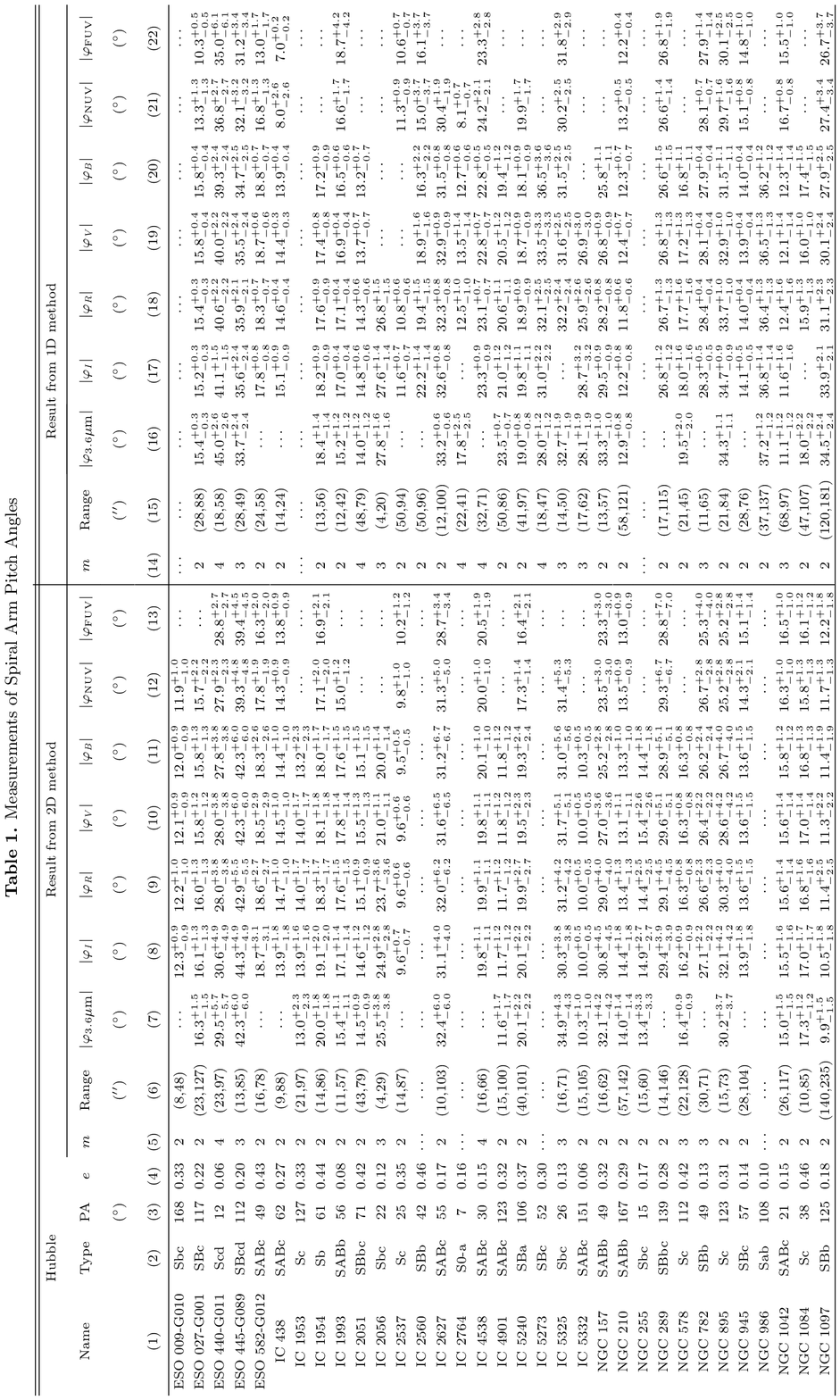}

\clearpage
\includegraphics{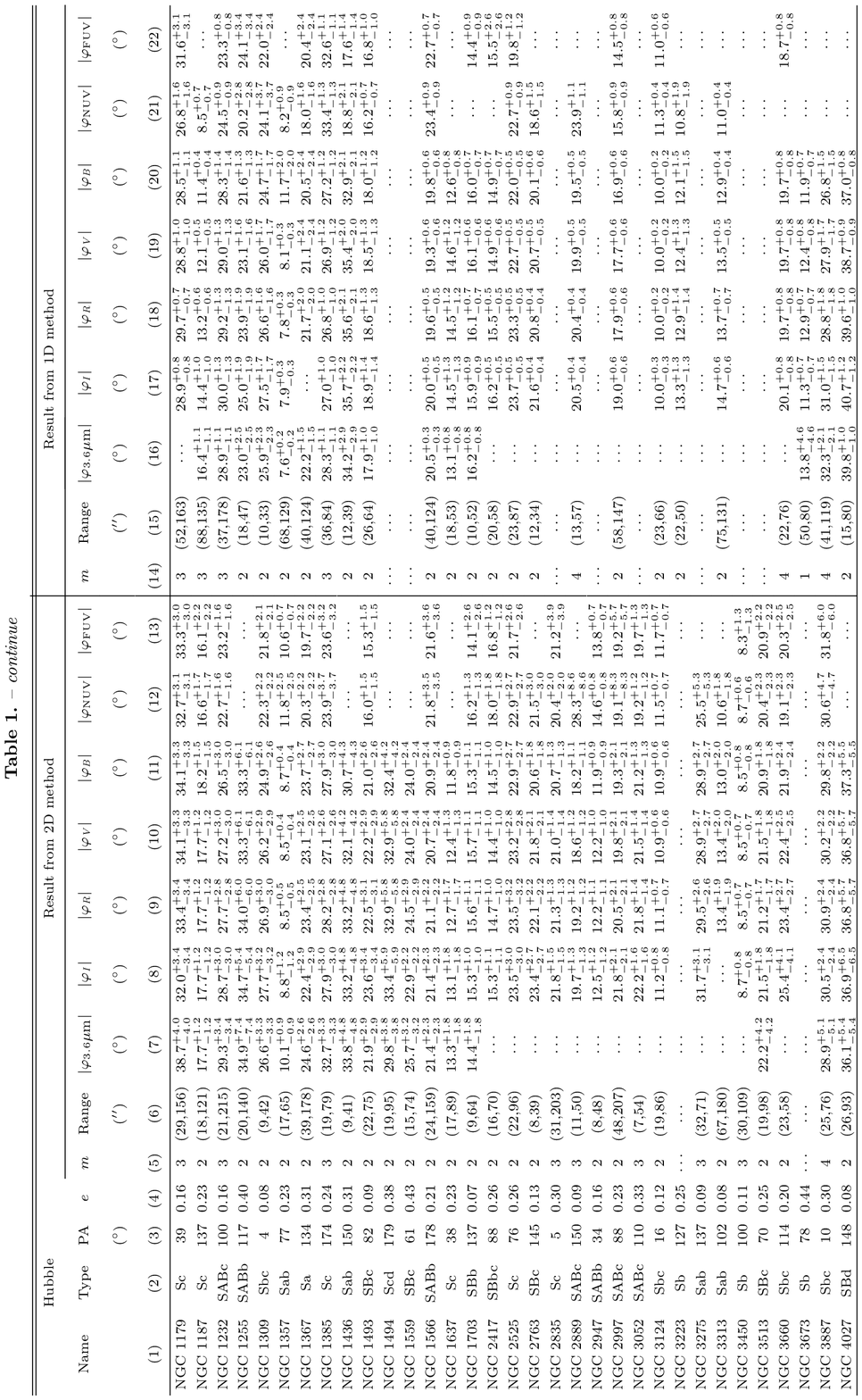}

\clearpage
\includegraphics{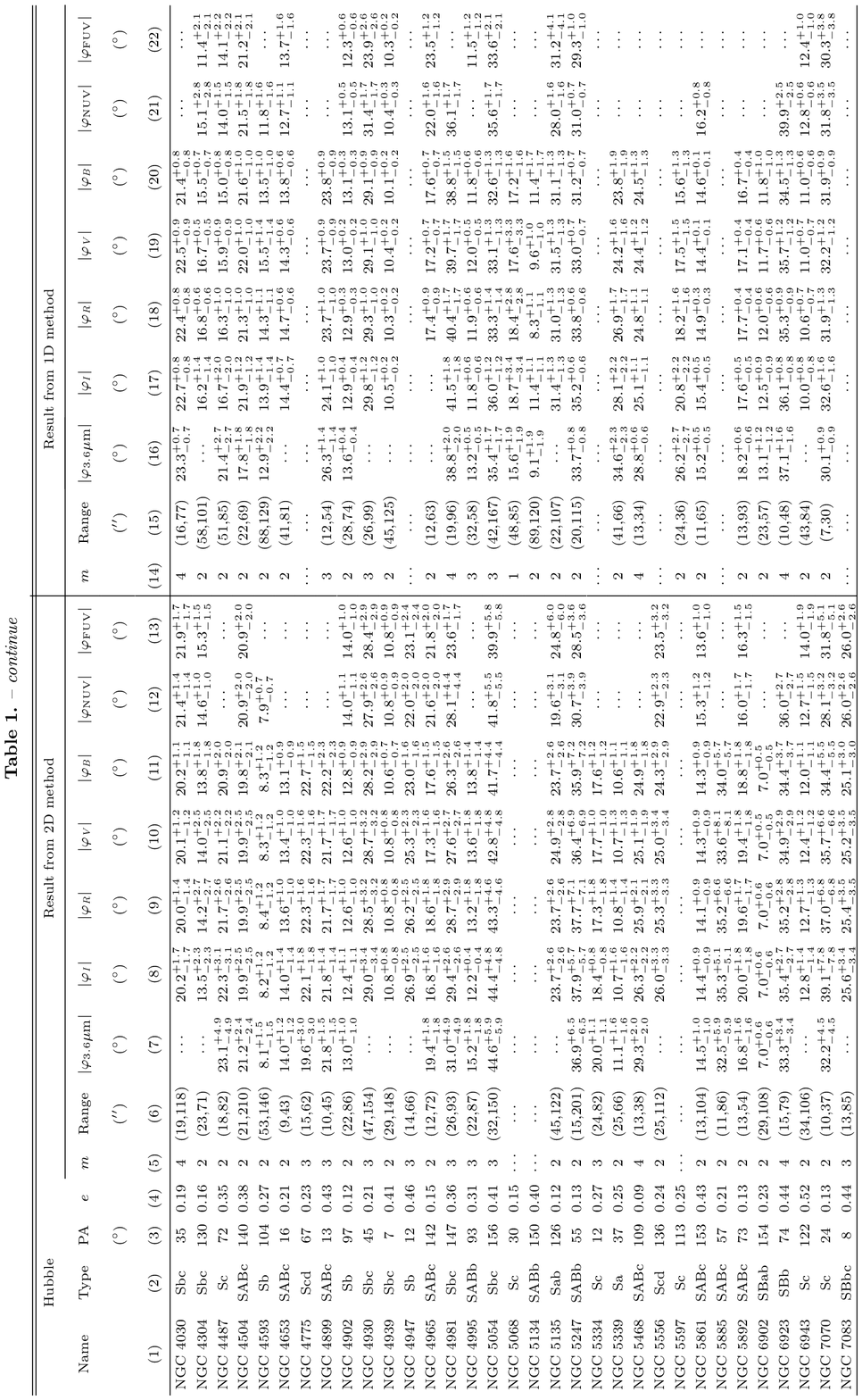}

\clearpage
\includegraphics{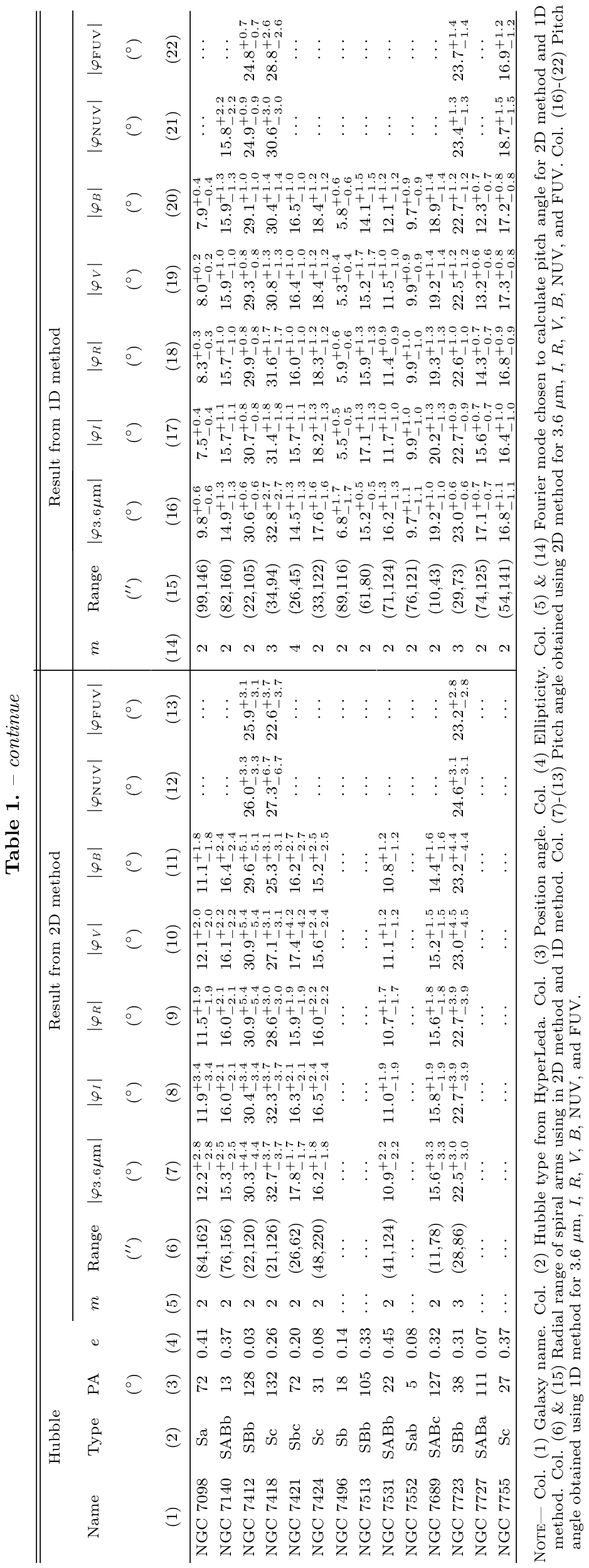}
\clearpage


\end{document}